\documentclass[%
 reprint,
 superscriptaddress,
 amsmath,amssymb,
 aps,
]{revtex4-2}

\usepackage{graphicx}% Include figure files
\usepackage{dcolumn}% Align table columns on decimal point
\usepackage{bm}% bold math
\begin{document}

%\preprint{APS/123-QED}

\title{Giant widening of interface magnetic \\ layer in almost compensated iron garnet} 
\author{Yu. B. Kudasov}
\email{yu\_kudasov@yahoo.com}
\affiliation{Sarov Physics and Technology Institute NRNU "MEPhI", 6, str. Dukhov, Sarov, 607186, Russia}
\affiliation{Russian Federal Nuclear Center - VNIIEF, 37, pr.Mira, Sarov, 607188, Russia}

\author{M. V. Logunov}
\affiliation
{Kotel’nikov Institute of Radio-Engineering and Electronics of RAS, 11-7 Mokhovaya Street, Moscow 125009}

\author{R. V. Kozabaranov}
\affiliation{Sarov Physics and Technology Institute NRNU "MEPhI", 6, str. Dukhov, Sarov, 607186, Russia}
\affiliation{Russian Federal Nuclear Center - VNIIEF, 37, pr.Mira, Sarov, 607188, Russia}

\author{I. V. Makarov}
\affiliation{Russian Federal Nuclear Center - VNIIEF, 37, pr.Mira, Sarov, 607188, Russia}

\author{V. V. Platonov}
\affiliation{Sarov Physics and Technology Institute NRNU "MEPhI", 6, str. Dukhov, Sarov, 607186, Russia}
\affiliation{Russian Federal Nuclear Center - VNIIEF, 37, pr.Mira, Sarov, 607188, Russia}

\author{O. M. Surdin}
\affiliation{Sarov Physics and Technology Institute NRNU "MEPhI", 6, str. Dukhov, Sarov, 607186, Russia}
\affiliation{Russian Federal Nuclear Center - VNIIEF, 37, pr.Mira, Sarov, 607188, Russia}

\author{D. A. Maslov}
\affiliation{Sarov Physics and Technology Institute NRNU "MEPhI", 6, str. Dukhov, Sarov, 607186, Russia}
\affiliation{Russian Federal Nuclear Center - VNIIEF, 37, pr.Mira, Sarov, 607188, Russia}

\author{A. S. Korshunov}
\affiliation{Russian Federal Nuclear Center - VNIIEF, 37, pr.Mira, Sarov, 607188, Russia}

\author{I. S. Strelkov}
\affiliation{Russian Federal Nuclear Center - VNIIEF, 37, pr.Mira, Sarov, 607188, Russia}

\author{A. I. Stognij}
\affiliation
{Scientific-Practical Materials Research Centre NAS of Belarus, 19 P. Brovki Street, Minsk 220072, Belarus}

\author{V. D. Selemir}
\affiliation{Sarov Physics and Technology Institute NRNU "MEPhI", 6, str. Dukhov, Sarov, 607186, Russia}
\affiliation{Russian Federal Nuclear Center - VNIIEF, 37, pr.Mira, Sarov, 607188, Russia}

\author{S. A. Nikitov}
\affiliation
{Kotel’nikov Institute of Radio-Engineering and Electronics of RAS, 11-7 Mokhovaya Street, Moscow 125009}

\date{\today}

\begin{abstract}
A two-sublattice ferrimagnet undergoes a transition from a collinear to canted magnetic phase at
magnetic field
oriented along an easy magnetization direction. In this work, we study the transition by means of the magneto-optical Faraday effect in a thin film of compensated iron garnet
(Lu$_{3-{\rm{x}}}$Bi$_{\rm{x}}$)(Fe$_{5-{\rm{y}}-{\rm{z}}}$Ga$_{\rm{y}}$Al$_{\rm{z}}$)O$_{12}$
grown on Gd$_3$Ga$_5$O$_{12}$ substrate. 
%The magneto-optical Faraday effect was used to determine the 
% magnetization state in the film under high magnetic field.
In the immediate vicinity of the compensation temperature a precursor of the transition with a complex shape was observed. Using a special sample with variable thickness we demonstrate
an interfacial origin of the precursor. 
Diffusion of gadolinium from the substrate into the film forms a thin intermixed layer with enhanced magnetization. It
induces an extended inhomogeneous magnetic structure in the film.
A two-step shape of the precursor appears due to an easy-plane anisotropy
of the intermixed magnetic layer. We emphasize that an effective width of the inhomogeneous magnetization distribution in the film grows enormously while approaching the compensation temperature.
\end{abstract}

\maketitle

Recently, antiferromagnets have drawn a wide attention as suitable magnetic materials to solve the problem of energy efficiency and performance of devices for information technology \cite{Jungwirth2016}. Antiferromagnets promise significant progress in this direction \cite{Jungwirth2016,Yang2017,Jungwirth2018,Fukami2020}. However, their advantages also give rise to technological difficulties in controlling and detecting antiferromagnetic (AFM) states because of the lack of pure magnetization. Ferrimagnets, e.g. iron garnets, at the point of magnetic compensation demonstrate a behavior similar to antiferromagnets. However, in contrast to them, properties of sublattices in ferrimagnets are not completely equivalent, even in the exactly compensated state. Therefore, it is possible to use well-known methods such as the magneto-optical Faraday and Kerr effects, the anomalous Hall effect and others to detect the state of compensated ferrimagnet \cite{Finley2020}.

Iron garnets (IG) crystallize in the cubic space group Ia3d and have the general chemical formula %%@
R$_3$Fe$_5$O$_{12}$, where R denotes a 
dodecahedral position usually occupied by rear-earth element \cite{garnets}.
In case of non-magnetic ion in this position a magnetic
behavior of IG is determined by Fe$^{3+}$ ($S=5/2$) ions in octahedral and tetrahedral
crystallographic positions which form two sublattices, correspondingly. Since quantities
of ions in the sublattices are in the ratio of $2:3$ and an interaction between the sublattices
is AFM, the magnetic structure turns out to be ferrimagnetic.
Prominent magneto-optical characteristics \cite{Zvezdin} and extremely weak damping of spin waves \cite{Deb} %%@
make
IG the basis of various optoelectronic and promising spin-wave devices development \cite{Zvezdin,Deb1,Dreyer}.

Another feature of these
compounds is flexibility of their composition \cite{garnets}, which allows adjusting smoothly magnetic and %%@
optical characteristics in a wide range.
For instance, a substitution of rear-earth ion R by bismuth is widely used for enhancement of magnetoopical %%@
properties \cite{Hansen}. 
Under specific technological conditions, a substitution of iron by non-magnetic gallium and aluminum
becomes selective, that is, up to 95~\% of substituents are located in the tetrahedral positions %%@
\cite{Roschmann}.
The selective dilution results in compensation of magnetic moments of the iron sublattices \cite{Hansen2}. In this case
a temperature dependence of the total magnetic moment $M(T)$ demonstrates a compensation point even if the %%@
dodecahedral sublattice is non-magnetic.
A strong disorder in the tetrahedral sublattice induced by the dilution leads to domination of the octahedral %%@
sublattice at low temperatures and tetrahedral one above the compensation point \cite{Kudasov1}.
Presently, an ultrafast magnetic switching effect in the compensated IG is intensively studied because of
great practical potential \cite{Shelukhin,Krivoruchko}.

The magneto-optical Faraday effect in IG was carefully studied for few decades \cite{garnets,Zvezdin}. 
A gap between crystal electric field levels for a diamagnetic-like dipole transition in Fe$^{3+}$ is defined by environment configuration.
That is why, the contributions of iron ions in octahedral
and tetrahedral positions differ and a magneto-optical response is observed even in
the compensated two-sublattice IG \cite{Deb2}. Although a mechanism of the drastic Faraday effect enhancement
with the substitution by non-magnetic bismuth ion is not obvious (e.g. see Ref.~\cite{Zenkov} and the %%@
following discussion Ref.~\cite{Helseth,Zenkov2}), one can practically use a simple model with the two independent diamagnetic transitions to describe the %%@
Faraday effect in Bi-substituted compensated IG \cite{Deb2}.

At the compensation temperature a magnetic behavior of the two-sublattice ferrimagnet is determined
by the difference between longitudinal and transverse
magnetic suseptibilities \cite{Zvezdin2}, i.e. by $k_u-h^2$, where $k_u$ is the constant of uniaxial
anisotropy and $h$ is the magnetic field. At a critical field the difference goes to zero and the transition %%@
to the canted magnetic structure occurs. Since the uniaxial anisotropy
and induced transverse anisotropy are cancelled out, the transformation of the magnetic structure is controlled %%@
by weak interactions and a variety of magnetic phase diagrams appears \cite{Zvezdin2}. 

Recently, the magneto-optical Faraday effect in thin film %%@
(Lu$_{3-{\rm{x}}}$Bi$_{\rm{x}}$)(Fe$_{5-{\rm{y}}-{\rm{z}}}$Ga$_{\rm{y}}$Al$_{\rm{z}}$)O$_{12}$  deposited on %%@
Gd$_3$Ga$_5$O$_{12}$ (GGG) substrate under high
magnetic field was investigated \cite{Kudasov3}. In the vicinity of
the compensation temperature, a precursor of the transition from the collinear
to canted magnetic structure was observed. The precursor had a step and subsequent plateau. 
One of possible explanations of this phenomenon was
an effect of interface magnetic layer which drastically broadens in the vicinity of the compensation point \cite{Kudasov2}.  

\begin{figure}
	\includegraphics[scale=0.44]{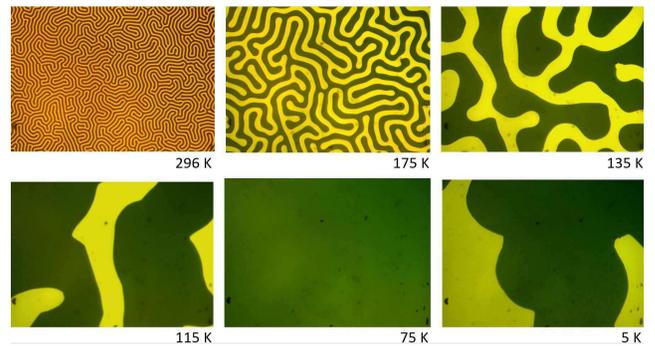}
	\caption{\label{f0} A domain structure in the film. The snapshots of the same area were taken in polarized light at different temperatures.}
\end{figure}

In the present Letter, we thoroughly investigate magnetic behavior of iron garnet film near the compensation point by means of precise Faraday rotation measurements at high magnetic field. The film (Lu$_{2.2}$Bi$_{0.8}$)(Fe$_{3.2}$Ga$_{1}$Al$_{0.8}$)O$_{12}$ was grown by liquid-phase epitaxy on a (111)-oriented single-crystalline GGG substrate \cite{Gerasimov}. The thicknesses of film and substrate were 9.1 and 500~$\mu$m, correspondingly. An easy magnetization axis was perpendicular to the film plane, the uniaxial anisotropy field was about 5~kOe. The sample was from the same batch as in Ref.~\cite{Kudasov3}. Due to the  bismuth substitution, the Faraday rotation at a wavelength of 633~nm, which was used in our experiments, was sizably enhanced and exceeds 8~deg. At the same time, the magnetic circular dichroism was less than 0.3~deg. Both of these factors were important for the measuring technique we applied. The film had the magnetic compensation temperature from 50 K to 78 K, according to preliminary magneto-optical measurements of the domain structure (Fig.~\ref{f0}, a characteristic domain size increased dramatically in the vicinity of the compensation temperature) and a temperature dependence of the transition field to the canted phase \cite{Kudasov3}.

Samples in the same batch had a dispersion of the compensation temperature because of slight composition variation over a plate. To exclude this effect, a square sample of about 1 cm$^2$ in area was cut out. A half of the sample was etched by low-energy ion-beam sputtering with an oxygen ion beam \cite{Logunov2019} down to 6.9~$\mu$m in thickness. Such a configuration of the sample allows us to perform measurements with films of different thicknesses but exactly the same composition to separate volume and surface contributions to the Faraday effect.

\begin{figure*}
	\includegraphics{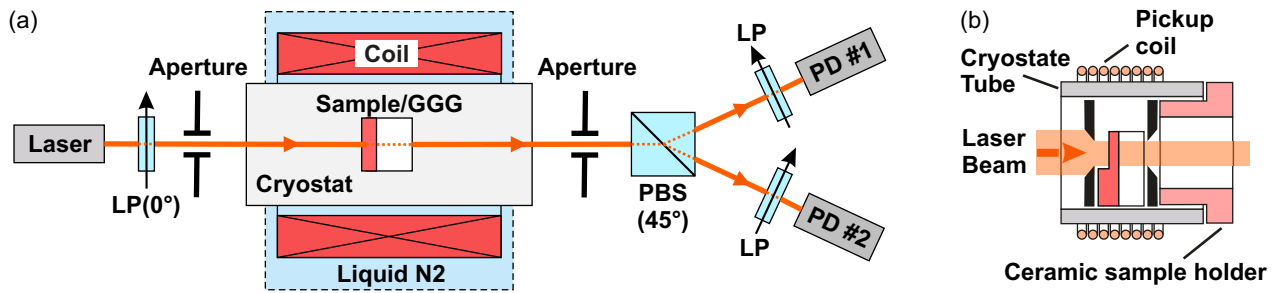}
	\caption{\label{f1} (a) A schematic representation of the optical path diagram, (b) an enlarged view of measuring unit with the sample.}
\end{figure*}

The sample was placed into an evacuated optical cryostat in the center of a pulsed 50-T solenoid %%@
\cite{Kudasov4, Platonov} (Fig.~\ref{f1}). A pick-up coil provided the measurement of
magnetic flux density in the sample with an accuracy of 3\%. 
Pulse rise and fall times of the magnetic field were about 2.5~ms and 10~ms \cite{Kudasov4}, correspondingly. 
The sample temperature was measured by a calibrated chromel-copel thermocouple within 1~K. An
additional low-power heater (not shown in Fig.~\ref{f1}) was placed at a sample holder to improve adjustment of temperature.

The Faraday rotation in the film was measured by modified technique developed earlier \cite{Kudasov3} (Fig.~\ref{f1}).
A beam from He-Ne laser ($\lambda=633$~nm)
passed through a polarizer LP(0$^0$) and thin or thick part of the sample, depending on a diaphragm position in the measuring unit as shown in Fig.~\ref{f1}b. Then it was split by the Wollaston prism (PBS)
into two beams %%@
with mutually orthogonal polarizations. 
Initially, the principal axes of prism were oriented at 45$^0$ to the incident beam polarization.
A pair of photodetectors (PD\#1 and PD\#2) produced signals, which were proportional to the light intensities in %%@
both channels ($I_1$ and $I_2$). The auxiliary polarizers (LP) were used for optimization of the light %%@
intensities at the photodetectors. The normalized difference $(I_1-I_2)/(I_1+I_2)$ allowed us to determine the %%@
angle of Faraday rotation \cite{Kudasov3}. The signal normalization effectively suppressed a drift and fluctuations of %%@
the laser intensity. The Faraday rotation angle was measured within 30${''}$.  

A contribution of the GGG substrate to the Faraday rotation was sizable and should be taken into account.
We have measured the Verdet constant of the GGG substrate in the same way as the sample and obtained the following values:
$V=13.3\pm0.5$~rad/(T~m) at 297~K and $V=14.7\pm0.5$~rad/(T~m) at 77~K. They are in a good agreement %%@
with the data of Ref.\cite{Haussuhl}.
It should be mentioned that the refinement of the Faraday rotation of the film by removing the substrate %%@
contribution leads to a progressive systematic error in the rotation angle of the following form:
$\Delta\varphi(B)=\alpha B$, where $B$ is the magnetic flux density and $\alpha$ is a constant %%@
($\leq$0.02~deg/T). Thus,
an insignificant slope uncertainty of $\varphi(B)$ dependence appears. 

\begin{figure}
	\includegraphics{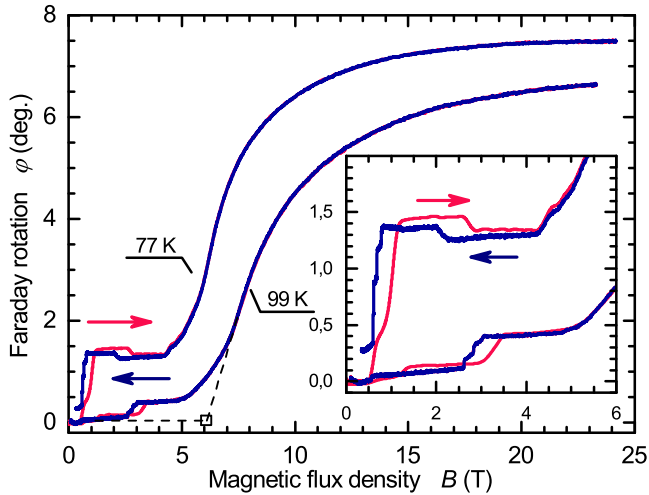}
	\caption{\label{f2} The Faraday rotation in rising (the red line) and decreasing (the blue line) magnetic %%@
		field at two temperatures. An enlarged initial part of the curves is shown in the insert.}
\end{figure}

The Faraday rotation in the thin part of the sample under increasing and decreasing magnetic field is shown
in the Fig.~\ref{f2}. Complex magnetic behavior was observed at low temperatures and magnetic fields below 6~T.
At high temperatures, the flat part of the curve at low magnetic field corresponded to the collinear magnetic structure. As the magnetic field increased, the curve character changed into steep ascent in the canted magnetic phase \cite{Kudasov3}. 
Irregular steps at the end of the falling branch of the magnetization curve appeared. They
are similar to those of observed in IG films, when domain structure formed under reversal
magnetization \cite{Hansen2}.  

\begin{figure}
	\includegraphics{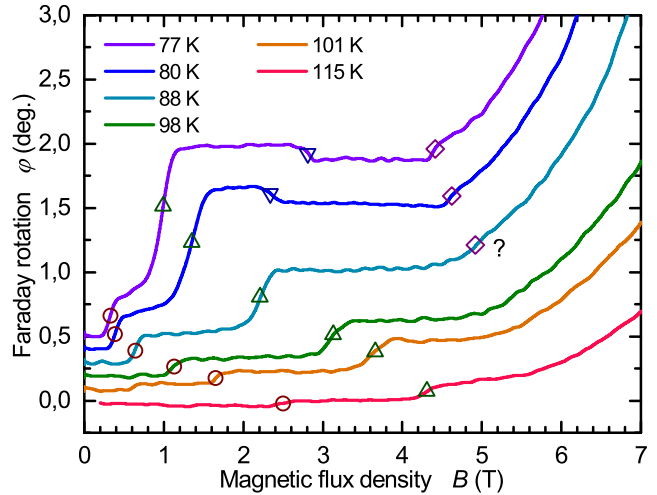}
	\caption{\label{f3} An evolution of the rising branch of the Faraday rotation with temperature. Specific points in the curve are marked with the symbols. The curves are vertically offset for clarity.}
\end{figure}

Figure~\ref{f3} shows an evolution of the Faraday rotation in the ascending branch with temperature.
There are two steps in the curves marked
as the up-triangles and circles. They grow with decreasing temperature, i.e. approaching
the compensation temperature. One of them was observed in the previous work \cite{Kudasov3}. Below 90~K,
a pair of small additional steps appeared (the down-triangle and diamonds). 

\begin{figure}
	\includegraphics{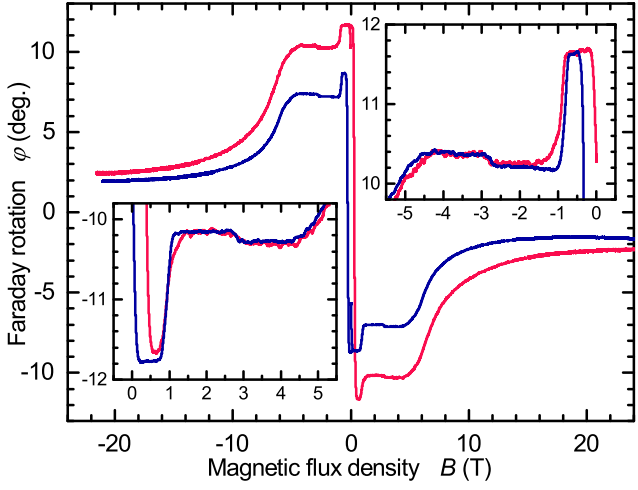}
	\caption{\label{f4} The Faraday rotation in the thick (the red line) and thin (the blue line) parts of the film. Enlarged fragments of the curves in the insert are shown with a vertical shift for the sake of %%@
		convenience.}
\end{figure}

A comparison of the Faraday rotation in the thick and thin parts of the sample is shown in Fig.~\ref{f4}.
The experiments were performed repeatedly with alteration of the magnetic field direction by
permuting solenoid leads. The hysteresis at small magnetic fields is related to film coercivity. The curves in Fig.~\ref{f4} allow evaluating a ratio of the rotation angles  in the thin ($\varphi_1$)
and thick $\varphi_2$ parts of the film, $\varphi_1/\varphi_2=0.74$. This value is in a good agreement with the interferometric measurements of the ratio by optical Vertex 80v spectrometer (0.76). 
In the inserts in Fig.~\ref{f4} enlarged views of low-field parts of the curves are presented.
A signal-to-noise ratio in the thin part of the film is much better because of larger light intensity at the %%@
sensors. It is clearly seen in the plot that heights of the steps in thin and thick parts of the film are the same, that is, they are independent of the film thickness.
This is an evidence of an interface origin of the low-field features in the film.

\begin{figure}
	\includegraphics{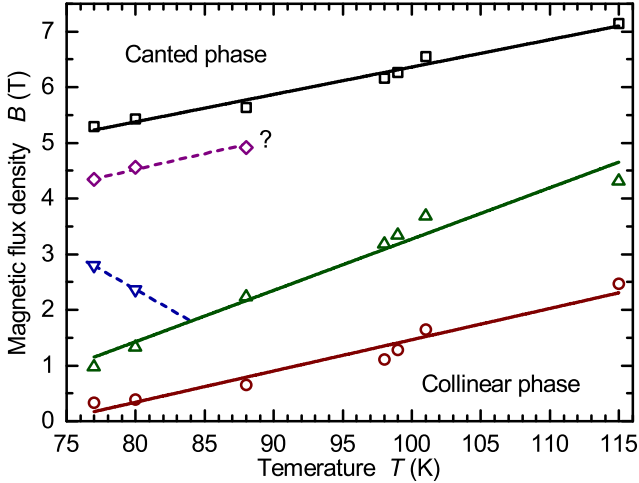}
	\caption{\label{f5} Phase diagram. The symbols correspond to the features marked in Figs.~\ref{f2} and \ref{f3}.}
\end{figure}

For the sake of convenience, the specific points of the curves in 
Fig.~\ref{f3} are represented in Fig.~\ref{f5} as a magnetic phase diagram. The bulk transition to the canted phase, which is defined as shown in Fig.~\ref{f2} by dash lines, is also depicted in the plot.

To understand the behavior of the magnetic interface layer let us firstly consider almost compensated %%@
ferrimagnet with uniaxial anisotropy. In case of our film, the easy axis is induced along the normal to %%@
substrate surface. Thus, the easy axis and external magnetic $h$ are oriented along the $z$ axis. Then the %%@
energy density of infinite ferrimagnet can be written down in the dimensionless form \cite{Zvezdin2}: 
\begin{eqnarray}
	f_v(\vartheta)=-h\eta\cos(\vartheta)-\left(k_u-h^2\right)\cos^2(\vartheta),
	\label{fv}
\end{eqnarray}
where $\vartheta$ is the angle between the AFM vector and $z$ axis, $k_u>0$ is the anisotropy constant,
and $\eta$ is the difference between magnetic moments of tetrahedral and octahedral sublattices. Moreover, it depends linearly on temperature, i.e. $\eta\propto T-T_c$. Here,
$T_c$ is the compensation temperature. 

Since the growth technique for the film involved high-temperature operations ($>$700~$^0$C), diffusion of gadolinium ions into the IG film with substitution of the ions in dodecahedral position took %%@
place \cite{Mitra}. A typical thickness of the intermixing depth is about 5~nm. In this layer the ions %%@
Gd$^{3+}$ ($S=7/2$) form the third magnetic sublattice coupled with the iron sublattices \cite{Mitra,Jakubisova}. 
This interaction leads to the surface energy  
\begin{eqnarray}
	F_0(\vartheta_0,\varphi)&=& g m \cos(\vartheta_0-\varphi) - h m \cos(\varphi) \nonumber\\
	& &+ m^2 k_s \cos^2(\varphi).
	\label{Fs0}
\end{eqnarray}
Here $g$ is the exchange constant between gadolinium and iron sublattices, $m$ is the magnetization value of %%@
gadolinium interface layer, $\varphi$ is the angle between the $z$ axis and gadolinium layer magnetization, %%@
$k_s$ is the anisotropy constant for the interface layer,
and $\vartheta_0 \equiv\vartheta(0)$, i.e. the value of the angle at the interface. The magnetization $m$ is described by %%@
the Brillouin function for $S=7/2$ at the compensation temperature. The magnetostatic energy of the magnetic interface %%@
layer is an obvious cause of an easy-plane anisotropy ($k_s>0$).
It should be mentioned that other sources of such an anisotropy at IG/GGG interfaces also exist \cite{Krichevtsov}.
The parameter $g$ can be estimated from a molecular field in Gd$_3$Fe$_5$O$_{12}$ \cite{garnets}. 
Assuming the dodecahedral sublattice to be half-filled by gadolinium, we obtain the molecular field of about %%@
35~T, that is, $g>>h$ and $g>>k_s$. Considering $h$ and $k_s$ to be small parameters, Eq.~(\ref{Fs0}) is %%@
reduced to the following form        
\begin{eqnarray}
	F_s(\vartheta_0)= &&- h \left( m_0 + g \chi \right) \cos(\vartheta_0) \nonumber\\
	& & 
	+ k_s \left( m^2_0 + g m_0  \chi \right) \cos^2(\vartheta_0),
	\label{Fs}
\end{eqnarray}
where $\chi={\rm{d}} m(h)\left/{\rm{d}} h\right. \left|_{h=g}\right.$ and $m_0=m(g)$.

We assume the surface layer to be much thinner than the film. That is why, we consider
semi-infinite ferrimagnet attached to the paramagnetic layer at the boundary. Then the total free energy of %%@
the inhomogeneous state can be represented by the following functional \cite{Kaganov}
\begin{eqnarray}
	\mathcal{F}\left[\vartheta\right]=\int_{0}^{\infty}\left(A \frac{\rm{d}^2\vartheta}{{\rm{d}}z^2} + %%@
	f_v[\vartheta] - f_v[\vartheta_{\infty}] \right) {\rm{d}} z + F_s(\vartheta_0),
	\label{FF}
\end{eqnarray}
where $\vartheta \equiv \vartheta(z)$ and $\vartheta_{\infty} \equiv \vartheta(\infty)$. The interface corresponds to $z=0$. The parameter $A$ is
of the order of squared interatomic spacing.
A variation of the functional gives the Euler-Lagrange equation, which in case of Eqs.~(\ref{FF}) and
(\ref{Fs}) is reduced to \cite{Kaganov}
\begin{eqnarray}
	A^{1/2}\frac{\rm{d}\vartheta}{{\rm{d}}z}= {\rm{sign}}(\vartheta_\infty - \vartheta_0) %%@
	\sqrt{f_v(\vartheta)-f_v(\vartheta_v)}
	\label{distr}
\end{eqnarray}
with the boundary condition at $z=0$
\begin{eqnarray}
	\frac{{\rm{d}}F_s}{{\rm{d}}\vartheta_0}= 2 A^{1/2} {\rm{sign}}(\vartheta_\infty - \vartheta_0) %%@
	\sqrt{f_v(\vartheta_0)-f_v(\vartheta_v)}.
	\label{bound}
\end{eqnarray}
Eqs.~(\ref{distr}) and (\ref{bound}) make up a system, which determines the function $\vartheta(z)$ and the %%@
free energy of inhomogeneous interface state \cite{Kaganov}. 
In contrast to Ref.~\cite{Kudasov2}, we have included into consideration the easy-plane anisotropy of
interface layer that is essential for further discussion. 

While the temperature tends to the compensation one ($\eta\rightarrow 0$), the solution of Eq.~(\ref{bound}) approaches asymptotically to an exact analytic expression. In case of the collinear bulk magnetic phase, i.e.$h<\sqrt{k_u}$, it takes a %%@
simple form:
\begin{eqnarray}
	\vartheta_0 = \arccos\left[\frac{2 A^{1/2} \sqrt{k_u-h^2}-h(m_0+g \chi)}{2m_0 k_s (m_0+g \chi)}\right].
	\label{exact}
\end{eqnarray}
Limiting values ($\vartheta_0 = 0$ and $\vartheta_0 = \pi$) give two critical fields.

\begin{figure}
	\includegraphics{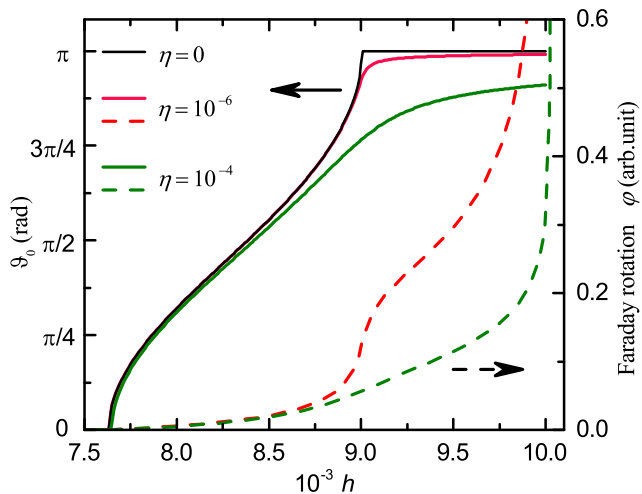}
	\caption{\label{f6} The calculated angle $\vartheta_0$ at different values of $\eta$ (the solid lines, left axis). The angle of Faraday rotation for the same $\eta$ is shown by the dash lines (the right axis). The black solid line is
		the exact solution Eq.~(\ref{exact}) for $\eta\rightarrow0$.}
\end{figure}

The results of numerical calculation of $\vartheta_0(h)$ at different values of $\eta$ are 
presented in Fig.~\ref{f6}  for the following set of parameters in Eq.\ref{Fs}: $m_0 + g \chi = 1.3\cdot 10^{-10}$ and  $ k_s \left( m^2_0 + g m_0  \chi \right) = 1.5 \cdot 10^{-13}$. These curves demonstrate
a temperature evolution of $\vartheta_0(h)$ because $\eta \propto (T-T_c)/T_c$. The curve $\eta=0$
corresponds to the exact solution Eq.~(\ref{exact}). 
The  angle of Faraday rotation estimated in the framework of the model of Ref.{\cite{Deb2,Kudasov2}}
is also shown in Fig.~\ref{f6}.

The shape of the $\vartheta_0(h)$ depicted in Fig.~~\ref{f6} has two step-like features. From the solution
Eq.~(\ref{exact}) one can see that it appears due to the easy-plane magnetic anisotropy
of the interface layer. With decreasing $k_s$, the two critical magnetic fields (at $\vartheta_0 = 0$ and $\vartheta_0 = \pi$) approach
each other and the function $\vartheta_0(h)$ reduces to a single step. Such a behavior in the absence
of the interface layer anisotropy was discussed in Ref.\cite{Kudasov2}.

It is worth discussing a width of the interface magnetic layer in the collinear phase, i.e. $h<\sqrt{k_u}$.
When $\vartheta<<1$, Eq.~(\ref{distr}) can be simplified to $\sqrt{f_v(\vartheta)-f_v(\vartheta_v)} \approx %%@
\sqrt{\eta h/2 +k_u - h^2} \vartheta$. Then $\vartheta(z)=\exp\left(-z/\delta\right)$, where
$\delta=A^{1/2}/\sqrt{\eta h/2 +k_u - h^2}$ is the width of the interface layer. When the bulk spin-flop
transition occurs, the width is $\delta=A^{1/2}/\sqrt{\eta h/2 }$. From here we see that the interface layer width diverges while approaching
the compensation temperature. That is why, the amplitude of
the steps marked by the up-triangles and circles in Fig.~\ref{f3} rises sharply.

The interface magnetic layer was studied earlier at IG/GGG boundaries by means of magneto-optical
Kerr effect \cite{Jakubisova}, polarized neutron reflectivity, and other techniques \cite{Mitra,Krichevtsov}.
The drastic growth of its width in the immediate vicinity
of the compensation temperature made possible its observation by the magneto-optical Faraday effect.
An estimation of the interface magnetic layer width from Fig.~\ref{f3} is about 700~nm. This is a huge value as compared %%@
with that of Yttrium IG film on the GGG substrate \cite{Jakubisova} of a nanometer thickness.

The model presented above also predicts the low-temperature magnetic behavior of the sample. Namely, the two large steps should disappear just below the compensation temperature. 

Additional small steps and plateau are observed in Fig.~\ref{f2} and Fig.~\ref{f3} at $T=$80~K and below (marked by down-triangles and diamonds). Their origin lies beyond
our present model and is still an open question. Most probably it is related to another free surface of the film.

%\section*{Acknowledgments}
%
The authors thank A. A. Fedorova, M. V. Gerasimov, and N. N. Loginov for technical support. The work was performed in the framework of National Center for Physics and Mathematics (Sarov, Russia). The work was partially supported by the Russian Foundation for Basic Research (Project No. 18-29-27020) and Russian Federation State support (Project No. 075-15-2019-1874).

\nocite{*}
\bibliography{LuBiFeGaAlO_bib_aps} 

\end{document}